\documentclass[10pt]{article}
\usepackage[OE]{express}
\usepackage{verbatim}
\usepackage{float}

\begin{document}

\title{Low-drift Zeeman shifted atomic frequency reference} 

\author{D.\,J.~Reed,\authormark{1,*} N. \v{S}ibali\'{c},\authormark{2} D.\, J. Whiting,\authormark{1} J.\ M. Kondo,\authormark{3} C.~S.~Adams,\authormark{1} and K.\,J.~Weatherill\authormark{1}}

\address{$^{1}$Department of Physics, Joint Quantum Centre (JQC) Durham-Newcastle, Rochester Building, Durham University, South Road, Durham, DH1 3LE, United Kingdom}
\address{$^{2}$Department of Physics and Astronomy, Aarhus University, Ny Munkegade 120, DK 8000, Aarhus, Denmark}
\address{$^{3}$Department of Physics, Federal University of Santa Catarina, Campus Trindade, 88040-900, Florian\'{o}polis, SC, Brazil}
\email{\authormark{*}dominic.j.reed@durham.ac.uk} 



\begin{abstract}
We present a simple method for producing a low-drift atomic frequency reference based upon the Zeeman effect. Our Zeeman Shifted Atomic Reference `ZSAR' is demonstrated to have tens of GHz tuning range, limited only by the strength of the applied field. ZSAR uses Doppler-free laser spectroscopy in a thermal vapor where the vapor is situated in a large, static and controllable magnetic field. We use a heated $^{85}$Rb vapor cell between a pair of position-adjustable permanent magnets capable of applying magnetic fields up to $\sim$1~T.
To demonstrate the frequency reference we use a spectral feature from the Zeeman shifted D1 line in $^{85}$Rb at 795~nm to stabilize a laser to the $7S_{1/2} \rightarrow 23P_{1/2}$ transition in atomic cesium, which is detuned by approximately 19~GHz from the unperturbed Rb transition. We place an upper bound on the stability of the technique by measuring a 2.5~MHz RMS frequency difference between the two spectral features over a 24~hour period. This versatile method could be adapted easily for use with other atomic species and the tuning range readily increased by applying larger magnetic fields.
\end{abstract}

\ocis{(020.7490) Zeeman effect; (020.2649) Strong field laser physics; (020.5780) Rydberg states; (140.3425) Laser stabilization; (140.3518) Lasers, frequency modulated.} 


\section{Introduction}
The frequency stabilization or `locking' of narrow linewidth lasers to atomic transitions is a common requirement in atomic and molecular physics experiments \cite{LaserReview1,LaserReview2,LaserReview3}. There are many well-established techniques for locking a laser on, or close to resonance including saturated absorption spectroscopy\cite{SpecReview1,SpecReview2}, polarisation spectroscopy\cite{PolSpec,PolSpec2}, dichroic atomic vapor level lock (DAVLL)\cite{DAVLL1,DAVLL2,AC_Coils}, modulation transfer spectroscopy \cite{ModTransfer1,ModTransfer3}, etc.
However, in many instances, it is desirable to stabilize a laser's frequency away from resonance. Methods to achieve this are typically more challenging. A common approach is to use a second laser which is referenced, or slaved, in some way to a laser that is stabilized to an atomic transition. Often this is achieved via a stabilized cavity\cite{HFCavity1,HFCavity2, ResInter}. Also, one can perform a beat measurement between the slave and reference lasers and stabilize the frequency difference between them\cite{BeatLocking1,BeatLocking2} or apply a sideband to the light using an electro-optic modulator (EOM) and seed the slave with the sideband\cite{SideBand1,SideBand2}. 
Often, it is desirable to stabilize a laser to a transition between two excited atomic states\cite{carr12}, in which case it is necessary to have additional lasers to first couple the ground state population to the lower of the excited states. An example of this is in the study of Rydberg systems where it is typical to use two \cite{EIT2} or more \cite{EIT3,EIT4} lasers. As transitions to Rydberg states typically have small electric dipole moments \cite{ARC}, large laser intensities can be required. 
An alternative approach is to use a transition in one atomic or molecular species to serve as a frequency reference to another species with a near-coincident transition energy. For example. iodine cells have long been used as a frequency reference in the infra-red due to the large number of well-characterized spectral lines in that range\cite{Iodine1,Iodine2}. There are also examples of one atomic species being used as a frequency reference for another in precision measurement experiments, for example in~\cite{Baird}.

Here, we introduce a simple method to provide a tunable frequency reference, based upon saturated absorption and frequency modulation spectroscopy\cite{FM1,FM2,FM3} of rubidium vapor in the presence of a large magnetic field. We demonstrate that the technique can provide a wide and continuously-tunable locking range in the vicinity of the D1 transition (795 nm). We use the technique to provide a frequency reference for the $7S_{1/2} \longrightarrow 23P_{1/2}$ transition in cesium which is approximately 19~GHz detuned from the unperturbed D1 transition in Rb \cite{RbMeas}. The method is essentially drift-free as it relies on large permanent magnets, and unlike methods using Faraday rotation~\cite{FaradayLock}, is insensitive to temperature fluctuations. Previously, in other work, demonstrations of using the Zeeman effect to provide a frequency offset have been made. E.g. Ref.~\cite{Armen} uses a 40~$\mu$m vapor cell \cite{ArmenCells} placed between permanent magnets to stabilize a laser to an atomic resonance shifted by $>$~5~GHz or Ref. \cite{ZeemanRbSpectr} wherein a stated application of their scheme is a stable reference setup. Because the Zeeman shifts are greater than the typical Doppler broadening, the method is also useful for simplifying atomic spectra into more cleanly defined systems and thus for precision measurements and quantum optics in thermal vapors\cite{DanEIA,DanEIT,Dan4WM,DanSP}.

\section{Modeling}

\begin{figure}[H]
	\centering
	\includegraphics[width=4.65in,height=3in]{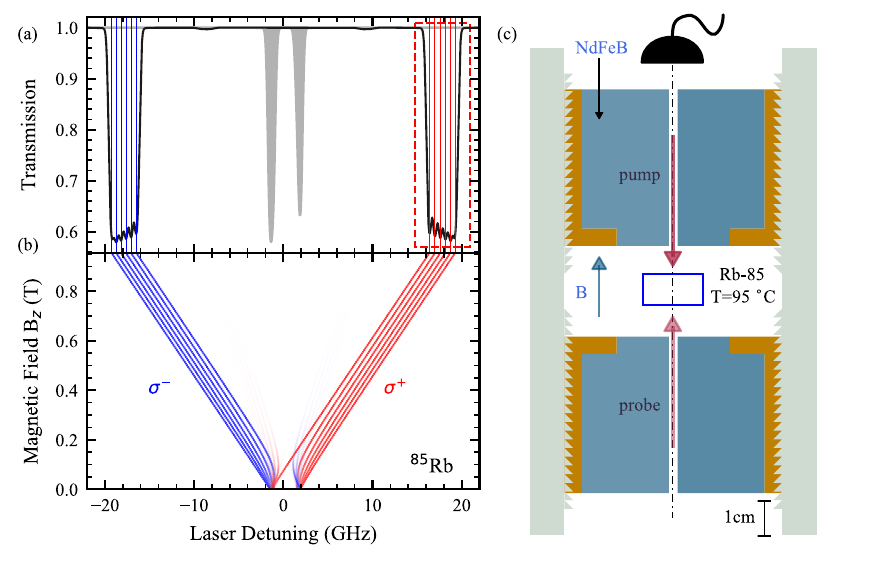}
	\caption[Rb85 B field evolution]{(a) A theoretical plot generated by ElecSus showing the transmission spectra of the D1 line of $^{85}$Rb at 0.954~T and 95~$^{\circ}$C as well as the same spectral line at zero field and 20~$^\circ$C displayed in gray for comparison. The region within the dashed box is used to provide the frequency reference in the experimental section. (b) A Breit-Rabi diagram showing the evolution of the spectral feature locations with increasing magnetic field, where $\sigma^{+}$ and $\sigma^{-}$ transitions are colored red and blue respectively. The opacity of each line indicates its transition strength, normalized to the strongest transition of this set. The small features at approximately $\pm$9.5~GHz are weak transitions that are allowed due to residual hyperfine mixing. (c) To scale diagram of the apparatus showing the B-field orientation relative to the probe and pump beam propagation directions. The experiment cell is a heated $^{85}$Rb cell with a 2~mm optical path length, flanked by NdFeB magnets.}
    \label{fig:B9540}
\end{figure}

The principle behind the locking technique is the Zeeman effect and so we give it the name `Zeeman Shifted Atomic Reference' (ZSAR). Figure~\ref{fig:B9540}(b) shows the evolution of the spectral feature locations of the $^{85}$Rb D1 line as a function of applied magnetic field strength where detuning is measured relative to the center of gravity of the hyperfine split ground and excited states. Also in Fig.~\ref{fig:B9540}(b), the opacity of each line indicates its respective transition strength, normalized to the strongest transition of this set. At high field, the atomic medium enters the hyperfine Paschen-Back regime, where the Zeeman interaction is greater than the hyperfine splitting. In this regime the total angular momentum is no longer a good quantum number\cite{Paschen}. Instead, the states are most easily described by the electron spin-orbit ($J$) and nuclear spin ($I$) angular momenta and their respective projections onto the magnetic field axis, $m_J$ and $m_I$. Using the ElecSus software~\cite{ElecSus}, which calculates the electric susceptibility of the medium, we are able to accurately predict the weak probe spectrum at arbitrary magnetic fields~\cite{FaradayEffect} and therefore infer the magnetic field strength at the position of the atoms from transmission measurements.  Figure~\ref{fig:B9540}(a) shows the Rb D1 spectrum calculated by ElecSus for a linearly polarised probe passing through a 2~mm vapor cell at 95~$^\circ$C in a magnetic field of 0.954~T. For comparison, the spectrum at zero field and 20~$^\circ$C is added in gray. The region inside the dashed box is compared to experimental data in section~\ref{sec:exp}. The vertical lines denote the frequency of $\Delta m_J = \pm 1$ transitions at high field shown in Fig.~\ref{fig:B9540}(a) where blue are $\sigma^{-}$ and red are $\sigma^{+}$ transitions, respectively. The very small features located at detunings of approximately $\pm$9.5~GHz are additional weak transitions that are allowed due to residual hyperfine mixing. This effect is taken into account in the full diagonalization, performed by the ElecSus software, of the atomic Hamiltonian in the presence of the magnetic field. More details of this process and the ElecSus software can be found in \cite{ZentileThesis}. There are no $\pi$ transitions as, in our model, the k vector of our probe beam and B field vector are parallel.

\section{Experiment}\label{sec:exp}

A  heated vapor cell with an optical path length of 2~mm containing $^{85}$Rb is placed between two magnets. To produce the desired magnetic field, we use a pair of custom designed NdFeB permanent magnets from first4magnets, mounted in a Helmholtz configuration with a separation of 9.7$\pm$0.1~mm. This separation was chosen as it was predicted to give a field strength of 9540~Gauss over the cell (our required field strength); which was later verified using a Hirst GM07 gaussmeter and by fitting weak probe transmission data using the ElecSus software. A diagram of the magnet apparatus is shown in Fig.~\ref{fig:B9540}(c). The cylindrically symmetric magnets each have an outer diameter of 49~mm and an inner diameter of 3~mm and individually produce an axial field of 0.5~T at the entrance to the bore. The inner bore holes allow clear passage of laser beams. The magnets also feature a small notch on their inward faces to facilitate secure mounting in a custom brace, which is threaded to allow precision adjustment of the magnet positions. In this way we can fine-tune the magnetic field strength in the region of the vapor cell. Detailed drawings and CAD files of the magnet holder setup for reproduction can be found at \cite{CAD}. 

\begin{figure}[H]
	\centering
	\includegraphics{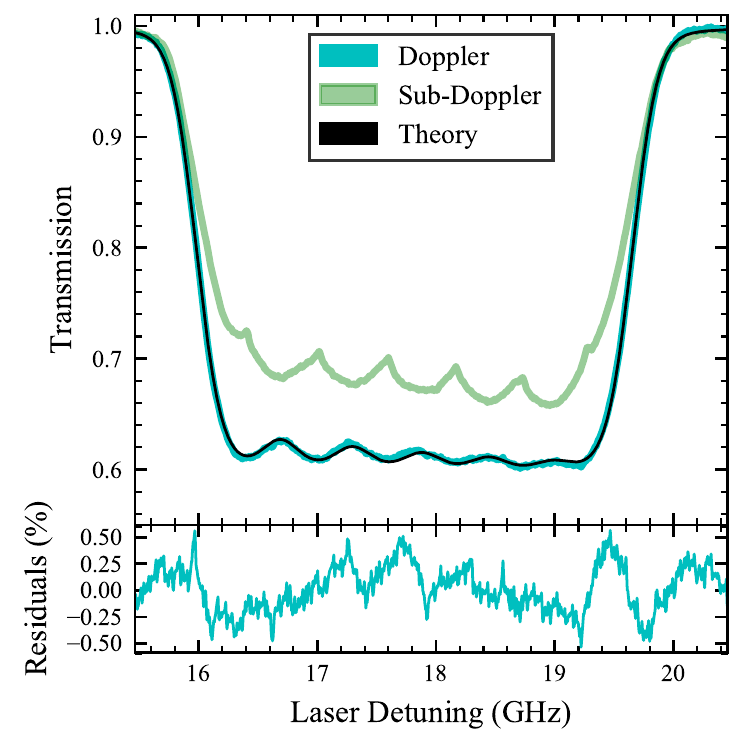}
	\caption{$^{85}$Rb Doppler broadened spectra which focuses on the positively detuned, $\sigma^{+}$ transition at B = 0.954~T with cell temperature of $\approx$ 95~$^{\circ}$C as denoted on Fig.~\ref{fig:B9540}(b) by the dashed box. The black line corresponds to an ElecSus theory plot using these experimental parameters, whilst the cyan line is the experimental Doppler spectra data. The residuals between theory and experiment are shown in the lower pane. The green line shows the sub-Doppler spectra of the Rb once a counter-propagating pump beam has been applied. It is from these resolved features that the FM signal is derived.}
	\label{Doppler}
\end{figure}

An external cavity diode laser operating at 795~nm is used to produce a weak `probe' beam and a strong `pump' beam. These beams, with 1/e$^2$ radii of 90$\pm10$~$\mu$m, counter-propagate and overlap within the vapor cell. In this geometry the beams are aligned with the magnetic field axis. The laser frequency scan is calibrated and linearized using a Fabry-Perot etalon. As the frequency of the laser is swept over the D1 transition, the probe beam transmission is recorded using a photodiode. At high magnetic field, the optical transitions separate and the optical density on resonance decreases. The temperature of the cell is stabilized at 95~$^{\circ}$C to maintain a sufficient optical density. Figure~\ref{Doppler} shows the experimental transmission of the horizontally polarised probe beam through the cell, with (green) and without (cyan) the counter-propagating pump beam. In order to calibrate the magnetic field strength and cell temperature, we initially use a weak probe, $I/I_{\rm sat}$~$\approx$~8$\times$10$^{-3}$\cite{WeakProbe}, to ensure good agreement with our theory curve (black) generated by ElecSus. This range of frequencies corresponds to the region in the dashed box in Fig. \ref{fig:B9540}(a). The agreement between theory and experiment is excellent when fitting to the experimental parameters of T = 96~$^{\circ}$C with an RMS error of 0.2~\%. The green plot shows the same probe transmission as before but now with a saturating pump beam (I/I$_{\rm sat} \approx 4$), counter-propagating the probe, as shown in Fig.~\ref{fig:B9540}(c). It is from this Doppler-free signal that our reference signal is derived as it resolves all of the individual transitions. There are no cross-over resonances in our spectrum as the features are further apart than the Doppler width. 

To generate a suitable dispersive-shaped reference for laser frequency stabilization, we use frequency modulation (FM) spectroscopy. To implement this, the probe beam is modulated using a 20~MHz EOM driven by an arbitrary function generator. The probe transmission, measured using a Hamamatsu C5460 20~MHz avalanche photodiode module, is amplified using a Minicircuits ZFL-500LN+ amplifier and mixed with the local-oscillator via a Minicircuits ZX05-1L-S+ mixer. The mixer output is then passed through a low-pass filter to remove the residual high-frequency components, leaving only the dispersive FM signal. Finally, the phase of the local oscillator is adjusted to maximize the FM signal amplitude.

\begin{figure}[H]
	\centering
	\includegraphics[width=4.725in,height=3.6in]{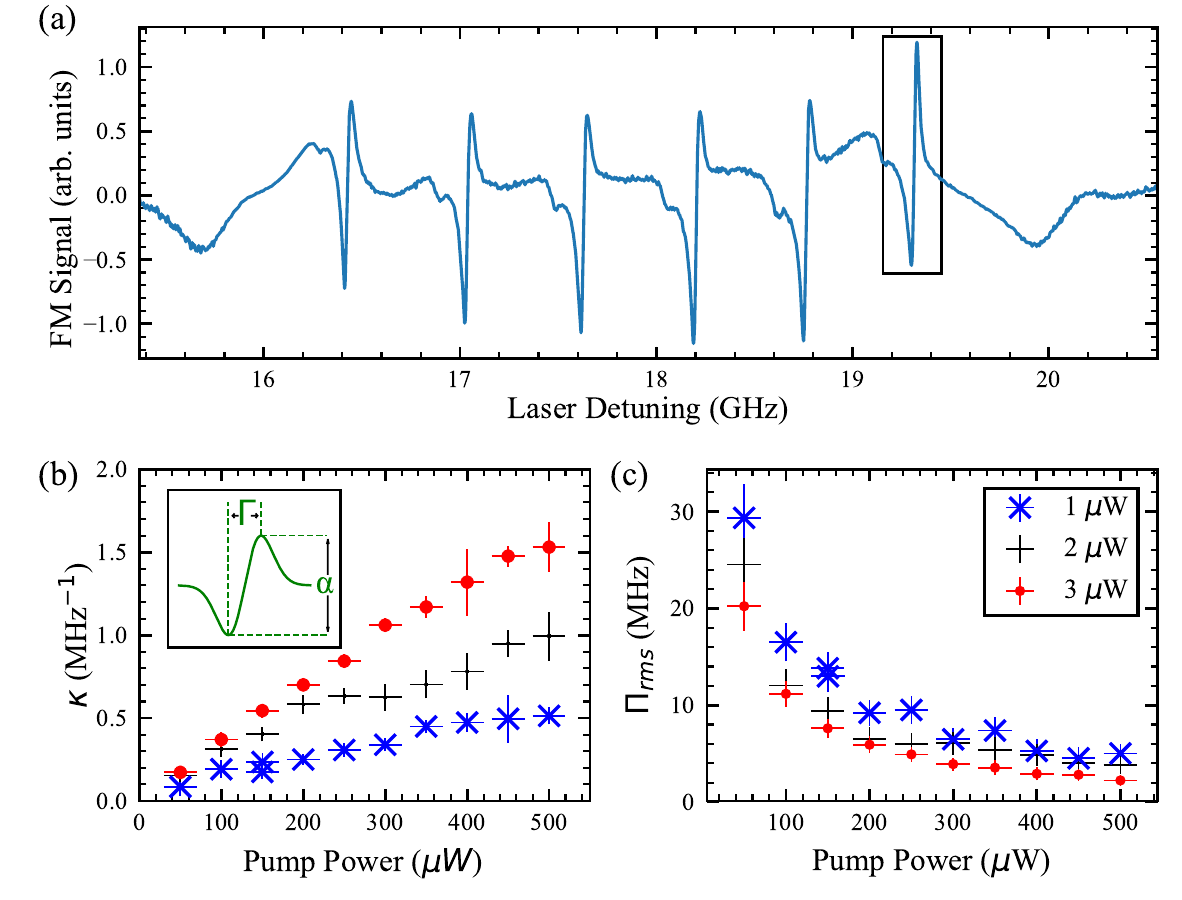}
	\caption[Laser setup]{(a)~Experimentally measured FM spectra showing the six dispersive features associated with the six sub-Doppler resonances appearing on the positively detuned part of the probe transmission spectrum. The feature highlighted by the black box was used to characterize our signal. (b) A study of the gradient, $\kappa$ of the FM feature contained within the black box as defined by $\alpha$, its peak to peak height, and $\Gamma$; its width. The data exhibits a linear relationship between $\kappa$ and the pump beam power. (c) An analysis of the uncertainty of the position of our locking feature $\Pi_{rms}$, in relation to both probe and pump powers. The minimum uncertainty was found to be 2.2$\pm$0.6~MHz.}
    \label{fig:FM}
\end{figure}

The measured FM spectrum is shown in Fig.~\ref{fig:FM}(a) wherein the six dispersive features correspond to the sub-Doppler resonances of the spectra in Fig.~\ref{Doppler}. For our locking signal characterization, we consider the maximally detuned feature (enclosed by the box). We begin by investigating the gradient $\kappa$ of the zero-crossing as we vary the pump and probe beam powers. We define $\kappa$ as the ratio of $\alpha$, the feature peak to peak value and $\Gamma$; its width. It is this quantity, along with the noise on the photodiode signal, which will primarily govern the characteristics of the laser frequency stabilization. Figure~\ref{fig:FM}(b), shows the evolution of $\kappa$ as a function of pump power for three different probe powers. There is a clear increase in the gradient with both pump and probe power over the range of parameters used. However, the power of the probe beam is necessarily kept below 5~$\mu$W to stay within the linear response range of the photodetector. We expect that this trend would eventually saturate due to power broadening and optical pumping.

Next, using the same data we can infer the RMS frequency uncertainty $\Pi_{\rm rms}$ of our zero-crossings by comparing the gradient at line center with the RMS noise, $\Sigma_{\rm rms}$ for each signal, I.e. $\Pi_{\rm rms}=\Sigma_{\rm rms}/\kappa$. $\Sigma_{\rm rms}$ was determined by taking the wings of each signal (which should be smooth in the absence of noise) and measuring their RMS deviations. The resulting frequency uncertainty can be seen in Fig.~\ref{fig:FM}(c); the minimum uncertainty was found to be 2.2$\pm$0.6~MHz in the case of 3~$\mu$W probe and 500~$\mu$W pump beams, respectively. Note that these were the highest pump and probe powers used and our data sampling rate for these measurements was 50~kHz. 

\begin{figure}[H]
	\centering
	\includegraphics[width=4.725in,height=3.6in]{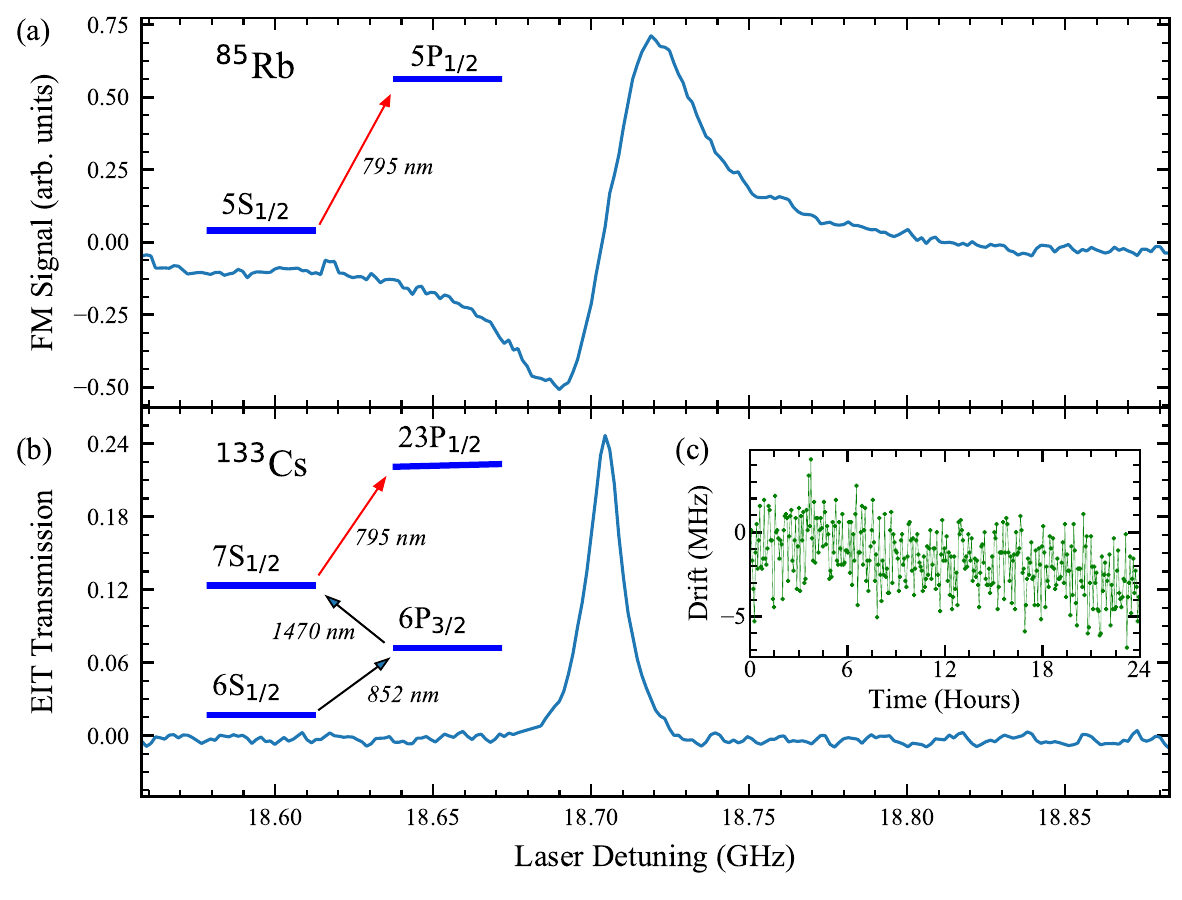}
	\caption[EIT]{(a) Frequency modulation (FM) signal generated using the 5S$_{1/2}$ $\longrightarrow$ 5P$_{1/2}$ transition in $^{85}$Rb at B = 0.954~T and (b) the simultaneous Rydberg EIT signal generated using Cs atoms on the $7S_{1/2} \longrightarrow 23P_{1/2}$ transition. Both plots are an average of 3 data sets. The FM signal probe and pump powers were 2~$\mu$W, and 300~$\mu$W respectively. Because of the addition of a double passed AOM to our EIT setup, the EIT signal is shifted by $\sim400$~MHz and therefore coincides with the 5th dispersive feature of the FM spectra. Characteristics of this alternate feature are consistent with our original analysis. (c) The measured long term drift in the relative positions of the EIT and FM spectral features, over a period of 24~hours. The maximum drift values were +4.3~MHz and -6.8~MHz with an RMS deviation of 2.5~MHz.}
    \label{fig:EIT}
\end{figure}

Finally, we place an upper bound on the drift of our method by measuring the resonant frequency of our dispersive feature with respect to the line center of another atomic resonance, generated independently. To that end we use a three-photon Rydberg EIT setup~\cite{EIT3} in cesium where the 7S$_{1/2}\longrightarrow$23P$_{1/2}$ transition is in close coincidence to the Rb D1 line. The level diagram is shown as an inset to Fig.~\ref{fig:EIT}(b) where the 6$^{2}$S$_{1/2}\longrightarrow$6$^{2}$P$_{3/2}$ transition is stabilized using polarisation spectroscopy and the 6$^{2}$P$_{3/2}\longrightarrow$7$^{2}$S$_{1/2}$ is stabilized via excited state polarisation spectroscopy \cite{PolSpec,carr12}. Figures~\ref{fig:EIT}(a) and~\ref{fig:EIT}(b) show the FM spectroscopy feature in Rb and the Rydberg EIT feature in Cs respectively. The spectra were recorded at the same time and compared over a 24--hour period. Figure~\ref{fig:EIT}(c) shows the relative drift between the two signals; the maximum recorded deviations were +4.3~MHz and -6.8~MHz with an RMS deviation of 2.5~MHz.

\section{Discussion}

The long term drift to negative frequency shown in Fig.~\ref{fig:EIT}(c) is dominated by the instability of the Rydberg EIT reference. This is caused by a drift in the reference voltage which provides an offset to the servo circuit used to stabilize the laser to the 6$^{2}$P$_{3/2}$ $\longrightarrow$7$^{2}$S$_{1/2}$ Cs transition. We therefore expect the drift in the ZSAR signal to be considerably less than the RMS drift measured here.

The short term RMS fluctuation in the reference frequency shown in Fig.~\ref{fig:EIT}(c) is consistent with fluctuations due to the RMS noise on the photodiode signal Fig.~\ref{fig:FM}(c). Thus, we can infer that ZSAR is highly robust to local perturbations as, in order to induce a shift of this magnitude, either the laboratory ambient field would have to fluctuate by $\pm$1.8~Gauss or the magnets themselves move by $\sim200~\mu$m. Both scenarios are highly unlikely. We therefore conclude that the limiting factor to our lock stability is the photodiode. If we were to replace it with a lower noise detector, we would achieve a greater signal-to-noise ratio and a reduced frequency uncertainty.

We note that the ZSAR method does not provide absolute frequency calibration but rather allows an arbitrary and stable offset to be applied to a well-defined atomic frequency reference. In  addition, the use of weak probe transmission measurements, such as those shown in Fig.~\ref{Doppler}, allows the magnetic field in the vicinity of the cell to be calibrated to approximately 0.5\% \cite{ElecSus}, similar to the precision of standard Hall probes.   


\section{Conclusion}

In conclusion, we have demonstrated a low-drift atomic frequency reference based upon the Zeeman effect with a large continuous, stable tuning range around the D1 line in rubidium. With the available field of $\approx$1~T we can achieve detunings of $\pm$20~GHz from line center of gravity with much greater detunings being possible; only limited by the applied magnetic field strength. We note that by comparing weak probe spectra to the ElecSus software it is not necessary for the laser to have a scan range that covers the entire spectrum.  

In direct comparison with the 3-photon EIT scheme for laser frequency stabilization\cite{EIT3}, the scheme presented here has clear advantages. ZSAR has a significantly simpler experimental setup, requiring only a single laser with around 1~mW of optical power. It is therefore cheaper and less sensitive to fluctuations in alignment and power than the EIT method. Additionally, the excellent passive stability and the large continuous tuning range provided by the permanent magnets allows for simple and robust on-resonant and off-resonant locking. 

Finally, our ZSAR scheme can trivially be extended to a greater tuning range by using stronger magnets and/or choosing another atomic transition. Eg. for the same magnetic field used here, by using the D2 line in $^{87}$Rb the detuning can be increased from 19.2~GHz to $\sim$26~GHz. This stable and versatile off-resonant frequency reference offers flexibility and simplicity for applications in cold atom and quantum optics experiments. 

\section*{Funding}
Engineering and Physical Sciences Research Council (EPSRC) grants EP/R002061/1,	EP/R000158/1, EP/M014398/1 and EP/M013103/1. Supported by DNRF through Niels Bohr Professorship.
\section*{Disclosures}
The authors declare that there are no conflicts of interest related to this article.
\section*{Acknowledgments}
The authors would like to thank James Keaveney for discussions and help with the ElecSus software template as well as Mingjiamei Zhang for her help with Matlab.\\

\end{document}